\documentclass[conference]{IEEEtran}
\IEEEoverridecommandlockouts
\usepackage{cite}
\usepackage{amsmath,amssymb,amsfonts}
\usepackage{algorithmic}
\usepackage{graphicx}
\usepackage{textcomp}
\usepackage{balance}
\usepackage{xcolor}
\def\BibTeX{{\rm B\kern-.05em{\sc i\kern-.025em b}\kern-.08em
    T\kern-.1667em\lower.7ex\hbox{E}\kern-.125emX}}

\makeatletter
\newcommand{\linebreakand}{%
  \end{@IEEEauthorhalign}
  \hfill\mbox{}\par
  \mbox{}\hfill\begin{@IEEEauthorhalign}
}
\makeatother

\begin{document}

\title{Group Authentication for Drone Swarms\\

}

\author{\IEEEauthorblockN{Yucel Aydin}
\IEEEauthorblockA{\textit{Informatics Institute} \\
\textit{Istanbul Technical University}\\
Istanbul, Turkey \\
aydinyuc@itu.edu.tr}
\and
\IEEEauthorblockN{Gunes Karabulut Kurt}
\IEEEauthorblockA{\textit{Department of Electrical Engineering} \\
\textit{Polytechnique Montréal}\\
Montréal, Canada \\
gunes.kurt@polymtl.ca}
\and
\IEEEauthorblockN{Enver Ozdemir}
\IEEEauthorblockA{\textit{Informatics Institute} \\
\textit{Istanbul Technical University}\\
Istanbul, Turkey \\
ozdemiren@itu.edu.tr}
\linebreakand 
\IEEEauthorblockN{Halim Yanikomeroglu}
\IEEEauthorblockA{\textit{Department of Systems and Computer Engineering} \\
\textit{Carleton University}\\
Ottawa, Canada \\
halim@sce.carleton.ca}
}

\maketitle

\begin{abstract}
In parallel with the advances of aerial networks, the use of drones is quickly included in daily activities. According to the characteristics of the operations to be carried out using the drones, the need for simultaneous use of one or more drones has arisen. The use of a drone swarm is preferred rather than the use of a single drone to complete activities such as secure crowd monitoring systems, cargo delivery. 

Due to the limited airtime of the drones, new members may be included in the swarm, or there may be a unification of two or more drone swarms when needed. Authentication of the new drone that will take its place in the drone swarm and the rapid mutual-verification of two different swarms of drones are some of the security issues in the swarm structures. In this study, group authentication-based solutions have been put forward to solve the identified security issues. The proposed methods and 5G new radio (NR) authentication methods were compared in terms of time and a significant time difference was obtained. According to the 5G NR standard, it takes $22$ \textrm{ms} for a user equipment (UE) to be verified by unified data management (UDM), while in the proposed method, this time varies according to the threshold value of the polynomial used and it is substantially lower than $22$ \textrm{ms} for most threshold values.

\end{abstract}


\section{Introduction}
The use of drones in military or commercial applications is enhancing. Search and rescue, cargo delivery, crowd monitoring, traffic monitoring, visual shows, threat detection in borders, and atmospheric research can be cited as examples of usage areas for drones \cite{crowdmonitoring,droneusage}. The capabilities of a single drone such as battery life, airtime, coverage area are limited to perform intensive tasks \cite{capabilities}. The use of a drone swarm consisting of more than one drone provides advantages over a single drone. Technological amenities presented to humanity always include some security issues. The study mostly focuses on the security aspect of drone swarms. 

\begin{figure}[h!]
\centering
\includegraphics[width=\linewidth]{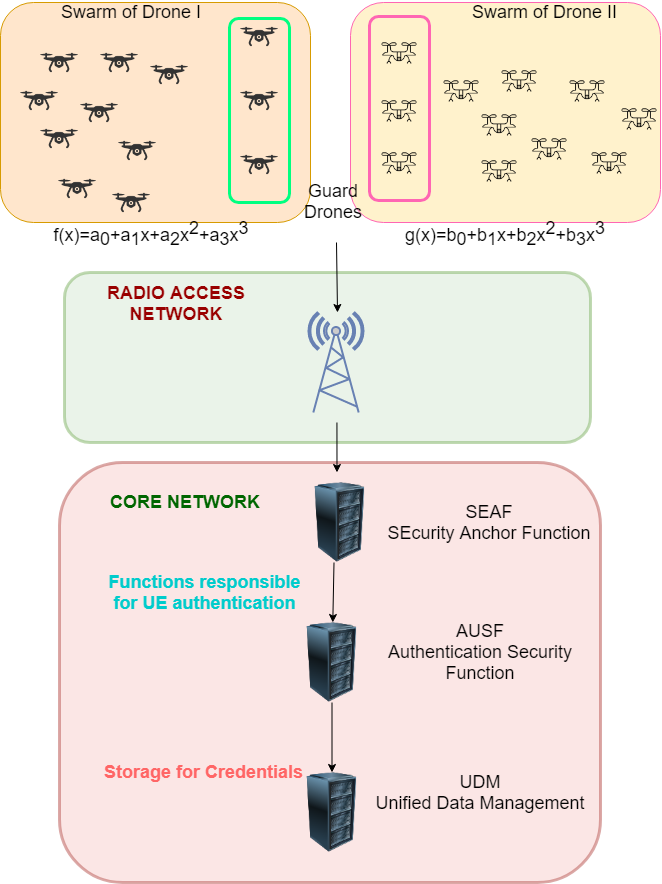}
 \caption{Unification of two drone swarms. \\ Two drone swarms can establish one larger swarm. Each member should authenticate the UAVs in the other swarm. If the requests for authentication are transmitted to the UDM for each member, scalability issues occurs. }
\label{fig:SwarmDrone}
\end{figure}

The study proposes solutions for two scenarios which are the authentication of new arriving drones and the unification of two separate swarms of drones. Including the new parties requesting to be part of the swarm without any authentication induces security issues such as intercepting the communication between drones, and routing the traffic to a malicious server. Our proposal for the authentication of the new parties is based on group-based solutions, which provides also a solution for latency problems. The reason for the latency issues in 5G new radio (NR) is requesting authentication from the core network for each new party. If the number of new parties is too much, the servers in the core network can cause latency. The guard drones as shown in Figure \ref{fig:SwarmDrone} can perform group authentication with the new parties and solve the latency problem by authentication more than one drone at the same time.

Due to the limitations mentioned before, deployment of new drones to the swarm or redeployment of any drone to the base takes place continuously. Even, the unification of two separate swarms of drones as shown in Figure \ref{fig:SwarmDrone} can be required if one drone swarm is not enough to accomplish the mission. Two swarms of drones may perform their tasks separately at the beginning. The control station controls the drones via cellular infrastructure. According to Release 17, authentication of each user equipment (UE) is accomplished through the core network functions such as security anchor function (SEAF), authentication server function (AUSF), unified data management (UDM), which is explained in detail in the next sections. 

In the next phases, there may be a requirement to unify two swarms of drones to perform intensive tasks. The authentication of each party in the two swarms one by one as mentioned in the 3GPP standards can cause scalability issues. 

These scenarios always come together with some security issues. The confidentiality and integrity of the communication between drones should be well secured. The new participant to the swarm should be authenticated and then should be included in the swarm.

With regard to the unification of two swarms of drones, the authentication of the members from the other swarm one-by-one as remarked in 3rd Generation Partnership Project (3GPP) 5G NR \cite{3GPP33501} standards is time and resource consuming. The authentication information should be transmitted to the base station from the authentication owner in the swarm and then to the core network for confirmation. A speedy authentication solution is required both for the new party and the unification.

This study is built on a group authentication solution in \cite{GA, GA2}. It is possible to authenticate more than one entity at the same by group authentication \cite{harn}. Guard drones are the parties in the drone swarm that are responsible for the authentication. Guard drones and new participants perform group authentication and if the new drone has valid credentials, communication with other members in the swarm is possible. Also, authentication between guard drones from the separate swarms of drones is enough for the unification of two or more swarms.

In light of these challenges, our main contributions are listed as;
\begin{itemize}

\item We propose a novel authentication scheme for a drone participating in a drone swarm with less time requirement than 3GPP Release 17. The credentials provided by the new party are not transmitted to the core network for verification. Guard drones and the new parties enforce group authentication to allow the new party to be part of the swarm.

\item It is possible to ensure a group key to the new party after testification of identity. The new party can communicate with other drones in the swarm with the group key.

\item The communication and computational complexity is too excessive if two separate groups trying to authenticate each other one by one as in the standards. It is possible to combine two or more drone swarms with less complexity in our proposed method. A subgroup of swarm called guard drones can perform authentication between each other on behalf of the complete swarm.

\end{itemize}

This paper is organized as follows. The next section provides an overview of the previous studies and authentication in the 3GPP standards. Our proposed approach for the authentication of new drones and unification of two swarms of drones in Section III. The performance evaluation is provided in Section IV and the study is completed by a conclusion in Section V.

\section{Related Works and Authentication in 3GPP Release-17}
The proposed method for the authentication of the new arriving drones is compared with the authentication schema in 3GPP standards. The details about the UE authentication in Release-17 and previous studies are explained in this section in order to remark the performance analysis section better.

\subsection{Related Works}
Previous studies on the security perspectives of drones or a drone swarm mostly focus on the attacks on the drones or the intrusion detection inside a drone swarm.

The security threats for the drones are examined in \cite{threat}. Drones are vulnerable to both physical and cyber-attacks. One of example for cyber attacks is mentioned hacker drones, which have capabilities to attack the other drones around them.

Blockchain technology is exploited in \cite{crowdmonitoring} to securely transfer data from a drone swarm to the cloud and identify the drones. Detecting abnormal activities in the crowd is one of the usage areas for the drone swarm.

In the study \cite{deeplearning}, the authors focused on the detection of drones performing abnormal activities in the drone swarms, neural networks are trained with normal behaviors and abnormal behaviors are tested.

In the study \cite{dronehack}, it is stated that the capture of drones by attackers does not only pose a security problem for the drone swarms or core network, but also drones controlled by malicious people threaten public safety.

A software-defined network-based solution is proposed in \cite{SDN} in order to improve the security of the drone swarm. 

The authors proposed a group key generation algorithm for a drone swarm in \cite{broadcastkey} and an authentication scheme for new joining drones. There is one manager drone and the other drones are followers. The proposed scheme depends on manager capabilities and availability. Most of the tasks are fulfilled by the manager drone. When a new drone participates in the swarm, the broadcast key is always updated by the manager. Moreover, the unavailability of one drone in the swarm blocks the regenerating of the broadcast key. 

The advantage of our proposed method over previous studies is to evaluate the authentication in the context of group authentication. Also, only guard drones inside the swarm are busy with the authentication processes. The other drones keep on to their own tasks during the authentication.

\subsection{UE Authentication}
The authentication of the new UE begins with the computing of subscription concealed identifier (SUCI) \cite{3GPP33501}, which is the encrypted version of the private key of UE. The private key of UE is the subscription-permanent identifier (SUPI) and the encryption key is the public key of a base station.

\begin{figure}[h!]
\centering
\includegraphics[width=\linewidth]{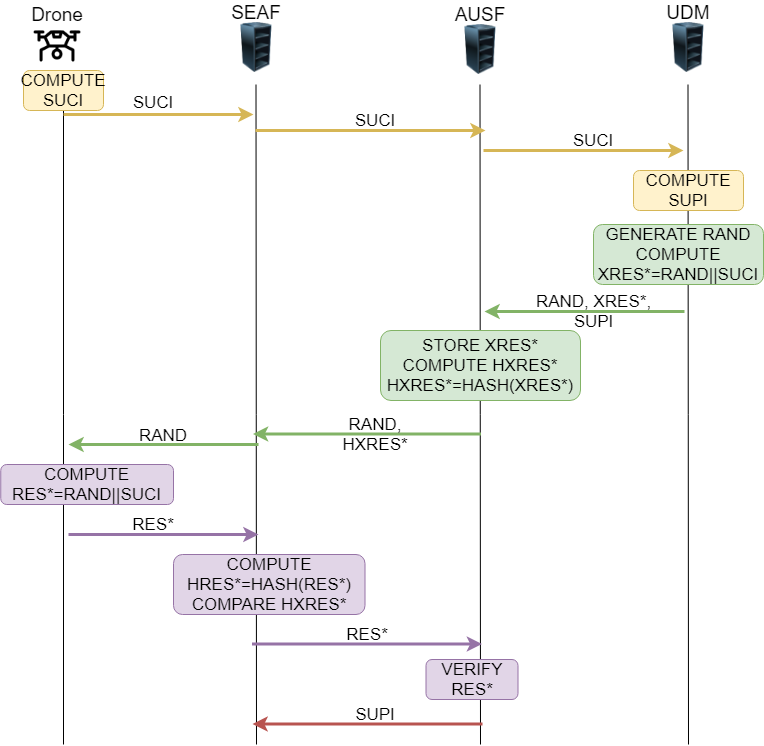}
 \caption{Authentication in Release-17. According to Release-17, the UEs should compute SUCI and send it to the UDM through SEAF and AUSF in the first phase of initial authentication. The UDM sends a challenge to the UE for confirmation by AUSF. }
\label{fig:authentication5g}
\end{figure}

The SUCI is transmitted to the UDM through the other functions as shown in Figure \ref{fig:authentication5g}. UDM decrypts the SUCI in order to extract the SUPI. A random is generated by UDM and appended with SUCI to have XRES. Random number, XRES, and SUPI are transmitted to the AUSF.

AUSF stores the XRES and computes the hash of XRES to obtain HXRES. AUSF sends HXRES and the random number to SEAF. The random number is transmitted to the UE. The UE computes RES by appending SUCI to the random number and sends it to the SEAF. SEAF can confirm the hash of RES by comparing it with HXRES. If the values are the same, SEAF sends RES to the AUSF. AUSF confirms the RES by comparing it with XRES. If the value is valid, AUSF sends SUPI to SEAF for upward secure communication.

\subsection{Security Issues for UE Authentication}
The UE authentication in 3GPP standards requires quite a few transmissions to verify the identity of UE by the core network. Apart from the scalability problem, the solution is also exposed to several attacks. The volume-based attacks \cite{volumeattack} are performed to consume the resources of legitimate elements of the mobile infrastructure such as BS, core network servers. If the authentication of the end devices is not designed well enough, the elements are exposed to attacks such as flooding attacks.

The authentication scheme in 5G NR depends on a long key stored in UDM and UE. The key is generated in the production phase of the USIM card. If the long key is captured by intruders, the whole scheme is collapsed \cite{USIMstealing}. The 5G authentication solution is exposed to SUPI-cracking attacks \cite{USIMstealing}. The intruder can encrypt random SUPI values by the public key of BS and send them to BS in order to strike valid SUPIs.

\section{Group Authentication Framework}

The proposed group authentication framework consists of two algorithms. The first algorithm can be exploited to authenticate new parties participating in the drone swarm.

\subsection{Inclusion of New Drones in the Swarm}
Drone swarms as a group are assumed to perform group authentication as indicated in \cite{GA}. The polynomial used for group authentication is in the form: 
\begin{equation}
f(x)=a_0+a_1x+...+a_{t-1}x^{t-1}.
\end{equation}
The value $a_0$ is the group key that enables the communication within the group via symmetric key encryption, and the $t$ value is the threshold value. Group authentication can be performed with up to $t$ valid public key pairs \cite{GA}. Therefore, the number of guard drones that will authenticate the new participant to the group should be $t-1$. If the new drone has a valid public key pair, the drone can perform group authentication with the guard drones as shown in Algorithm 1.

The authentication steps are listed as:

\begin{enumerate}
	\item The new drone shares its public key pairs ($x_{drone}, f(x_{drone}) \cdot P$) with guard drones ($P$ is a predefined elliptic curve point on a curve).
	\item Each guard drone shares its public key pairs $$(x_i, f(x_i) \cdot P)$$ with other guard drones ($i$ indicates the number of guard drone). 	
	\item Guard drones compute 
		\begin{equation}
		c_i=f(x_i)\times P{\overset{t}{\underset{r=1, r\neq i}{{\displaystyle\prod}}}(-x_r/(x_i-x_r))}
		\end{equation}
 	for each guard drone.
	\item If ${\overset{t}{\underset{i=1}{{\displaystyle\sum}}}c_i}$  is equal to $Q=a_0 \cdot P$, the new drone is a legitimate drone.
\end{enumerate}

After the new group member is authenticated, the group key ($f(0)$) is encrypted by any guard drone and shared with the new drone. The encryption key can be computed using the elliptic curve Diffie-Hellman key exchange scheme \cite{GA, GA2}.

\subsection{Unification of Two Drone Swarms}
In addition to the inclusion of the new drone in the drone swarm, the need for combining two swarms of drones may arise. An example would be the merger of two swarms of drones delivering cargo to two different regions and continuing to deliver cargo to a larger area. Each drone swarm uses a different private polynomial ($f(x), g(x)$) for group authentication as shown in Figure \ref{fig:SwarmDrone}. The polynomials are known only to the core network. Each drone within the groups has only public and private key, which are ($x_{drone}, f(x_{drone}) \cdot P$) and ($f(x_{drone})$) for group I and ($x_{drone}, g(x_{drone}) \cdot P$) and ($g(x_{drone})$) for group II, respectfully.

Algorithm 2 can be used by two groups to authenticate each other and to determine the group key for the new group created.

The steps are listed as:

\begin{enumerate}
	\item A guard drone determined by the guard drones in two groups requests private and public key pairs from the core network for the opposing group.
	\item The core network shares valid keys with the respective guard drone by encrypting the keys with the private key of the guard drone.
	\item The guard drone shares the public key pairs with the opposite group after decrypting the keys.
	\item Guard drones in the opposing group send each other their public key pairs.
	\item As each drone in the opposite group obtains a valid public key equal to the threshold value, it performs group authentication.
	\item If the group authentication is valid, the group key of the second group is encrypted and sent to the sharing guard drone in the first group.
	\item The guard drone in the first group encrypts the second group's key with the first group's group key and sends it to all group members.
	\item After this stage, two groups are combined and the group key of the second group is used in intra-group communication.
\end{enumerate}

\begin{figure}[h!]
\centering
\includegraphics[width=\linewidth]{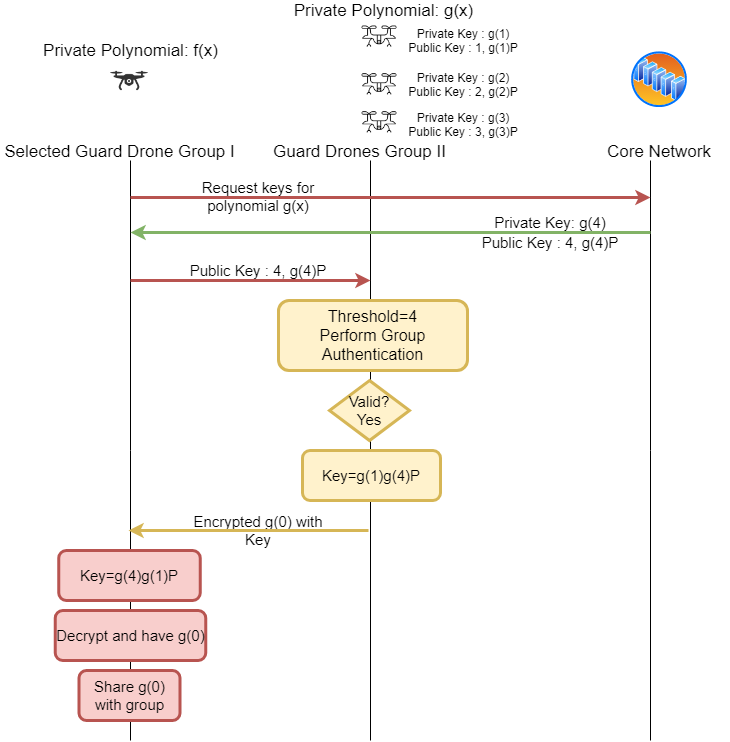}
 \caption{Authentication of two drone swarms. When the threshold value for the scheme is selected $4$, one drone from the first swarm and three guard drones from the second swarm are enough to perform group authentication.}
\label{fig:twoswarm}
\end{figure}

An example of two swarms of drones confirming each other is shown in Figure \ref{fig:twoswarm}. In the example, the threshold value is four, and three drones are the guard drones of the second group, and one drone is the guard drone selected from the first group.

\section{Performance Analysis}
The proposed scheme and the UE authentication method used in 5G NR \cite{3GPP33501} standards were compared in order to analyze the performance of the study.

To authenticate the new drone joining the drone swarm, either the proposed method can be used or the UE's private key ( SUPI in 5G NR) is sent to the core network for new UE authentication as specified in Release 17. The key is confirmed via a query in the database by the core network servers and the new drone is started to be served. In our proposed method, guard drones authenticate the new drone with the group authentication method.

The end result in the performance analysis is to find the time required to verify the identity of the new drone included in the group. This period will be found using both the proposed method and the method in the standards. The simulation is implemented by SimuLTE version 1.2.0 \cite{simulte} library built on top of the Omnet++ package version 5.6.2 and INET framework version 4.2.2. One UE, one BS, and core network servers are deployed to the simulation in order to monitor the time required for packet transmission. The simulation parameters in the omnet.ini file are left as default. These simulation settings are configured to have the authentication time for the standards. WiFi direct module in SimuLTE is exploited to monitor the time for the communication between UAVs. Two wireless hosts are deployed and a message packet from one host to other is sent to monitor the time. The communication for the authentication in the proposed method is mostly between UAVs. Therefore, the second configuration is about the proposed method.

\begin{figure}[h!]
\centering
\includegraphics[width=\linewidth]{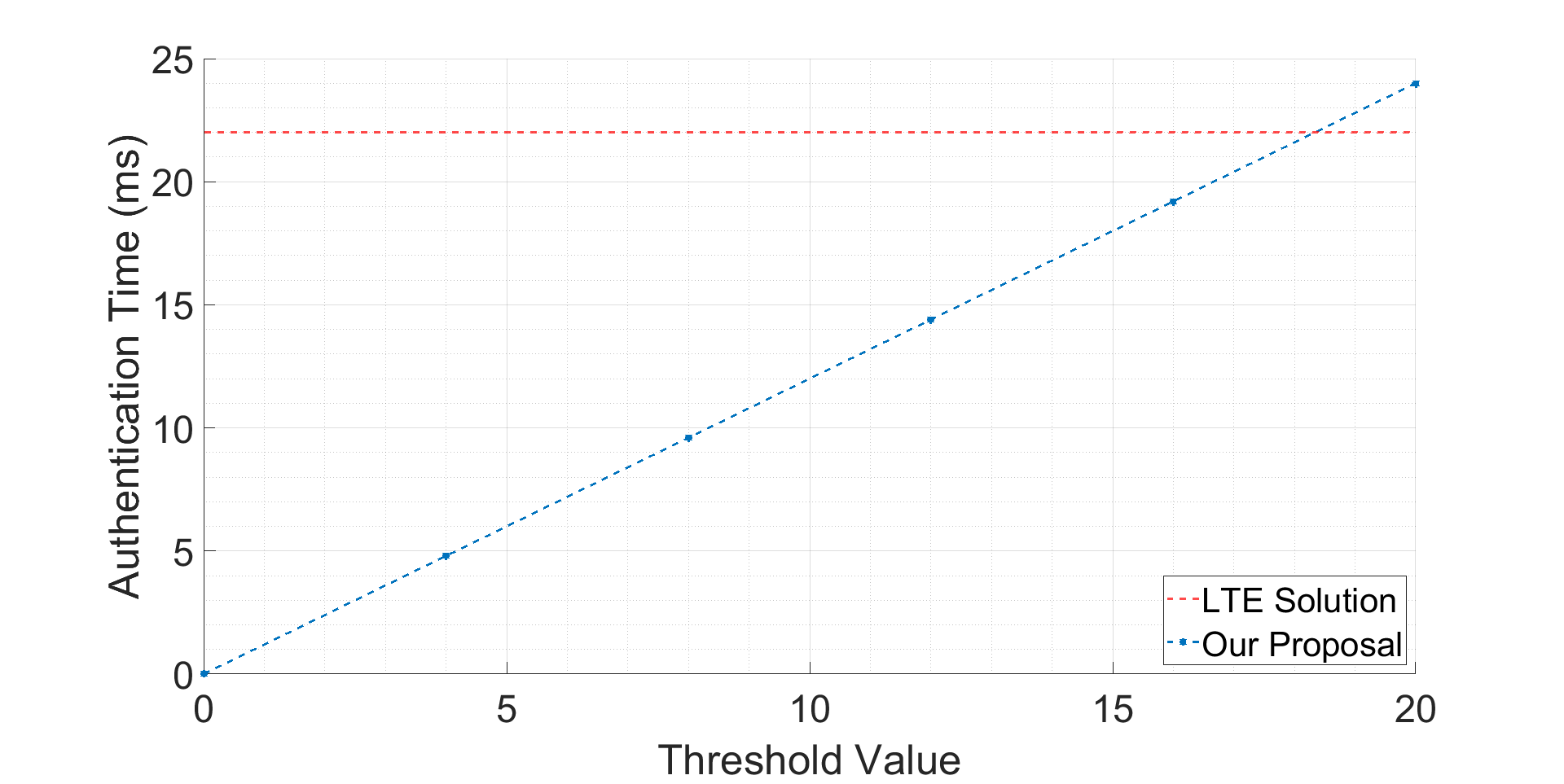}
 \caption{Comparison of authentication time in 5G NR and our proposal. Whenever the most significant power of polynomial is less than $10$, the proposed method ensures preferable time complexity than 5G NR.}
\label{fig:performance}
\end{figure}

It has been observed that the time required for a data packet sent by the UE to reach the server in the core network and for the server to send the response back to the UE is approximately $10$ \textrm{ms}. The UE performs one asymmetric encryption operation, which is $100$ \textrm{$\mu$s} \cite{ECP}, to compute SUCI and then sends the SUCI to the core network. The UDM decrypts the SUCI to have SUPI for confirmation. The time for the decryption is approximately $1.5$ \textrm{ms}. Two hashing operations are carried out in the scheme. The UE sends SUCI first and then RES to the core network for affirmation. The time required for the method in the standards is $22$ \textrm{ms} in total.

Data exchange time between drones was measured to simulate the proposed method. The wireless direct was carried out with SimuLTE and it was observed that the data transfer took place within $600$ \textrm{$\mu$s} between drones. Each guard drone shares data with the drone as much as the threshold value. Each guard drone also performs an elliptic curve powering operation up to the threshold value. One elliptic curve powering operation is approximately $612$ \textrm{$\mu$s} \cite{ECP}. In total, new drone authentication is performed in approximately $1.2t$ \textrm{ms} ($t$ is the threshold for polynomial) as shown in Figure \ref{fig:performance}. 

The proposed method provides faster authentication solution than 3GPP Release-17 according to the simulation results if the most significant power of the polynomial is chosen less than $10$.
The UAV should communicate with the core network two times for the initial authentication according to Release-16 and Release-17. 

\begin{figure}[h!]
\centering
\includegraphics[width=\linewidth]{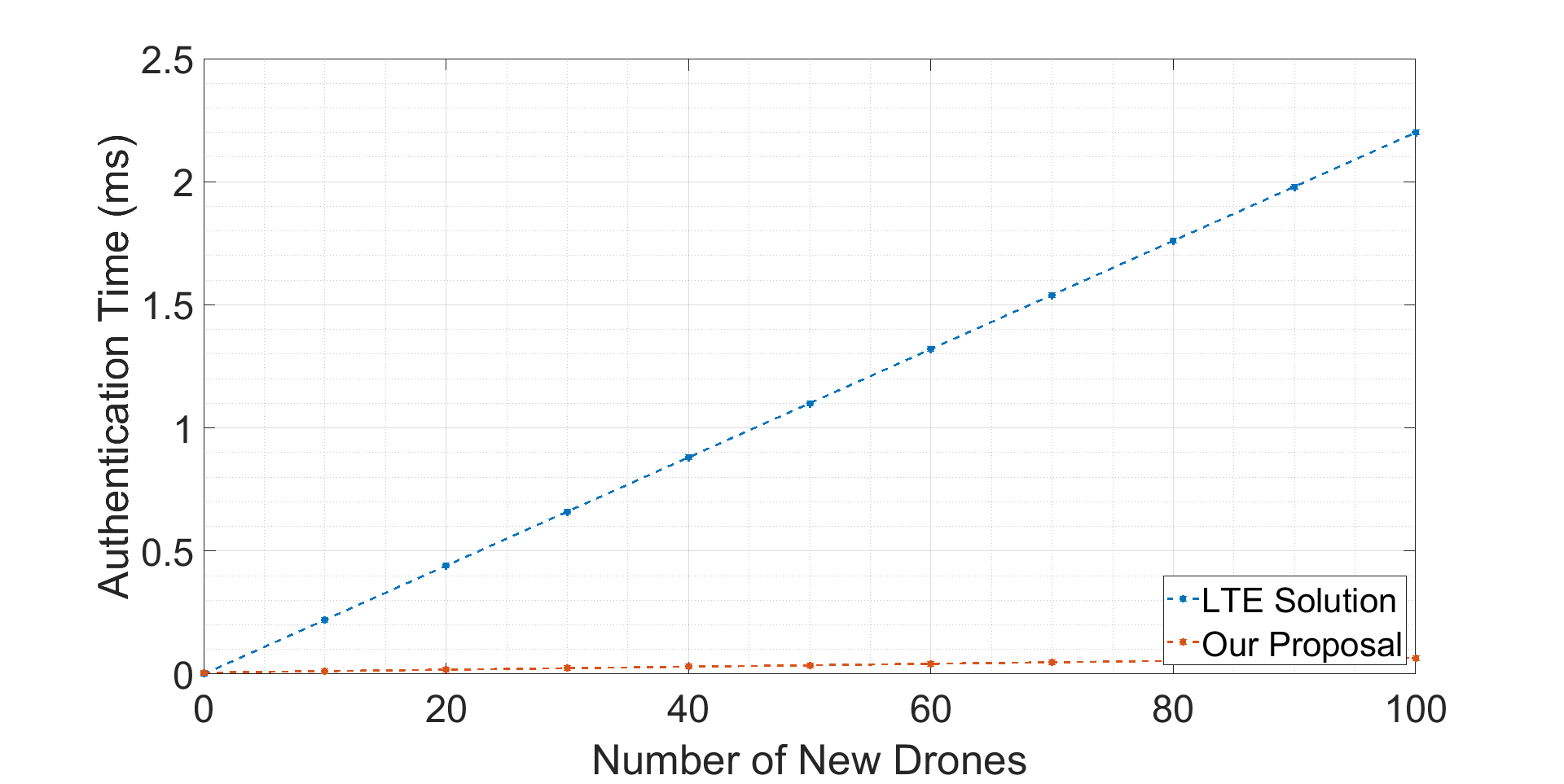}
 \caption{Authentication time for diverse number of drones. With increasing the number of drones, the time differences between the proposed method and 5G NR is changing swiftly.}
\label{fig:performance2}
\end{figure}

If it is assumed that $100$ new drones want to participate in the drone swarm, the time required by the core network for authentication is approximately $2.2$ \textrm{s} ($22$ \textrm{ms} x $100$ drones) for the 5G NR method as shown in Figure \ref{fig:performance2}. With the proposed method, the same number of drones can be authenticated as a group. Each party should broadcast its public key pairs, which cost $600$ \textrm{ms} ($600$ \textrm{$\mu$s} x $100$ drones), and one group authentication should be performed, which cost $6$ \textrm{ms} if the threshold value is selected as $5$. Total authentication time is almost $0.07$ \textrm{s}, which is extremely less than the 5G NR solution. According to International Mobile Telecommunications-2020 requirements, 5G NR should provide $20$ \textrm{Gbps} data rate and $1$ \textrm{ms} latency \cite{IMT}.

\section{Security Analysis}
In the security analysis section, the security provided by the proposed methods for all possible attacks are explained one by one. The sample for an attack is given as an attack and the solution is proved in the prevention part for each attack scenario.
\\

\textbf{Attack 1:} \textit{An intruder can eavesdrop on the traffic between guard drones and a new party and then perform a replay attack on the guard drones.}

\textit{Prevention.} In the proposed method, a key agreement phase is performed between the guard drone and the new drone in order to share the group key. The new drone must have a private key ($f(x_i)$) to decrypt the message coming from the guard drone.

\textbf{Attack 2:} \textit{In the unification of two drone swarms, an eavesdropper may capture the private key for the new participating drone.}

\textit{Prevention.} The intruders can capture the public key $(x_i,f(x_i)P)$ of the new drone by sniffing, but it is not possible to the private key $f(x_i)$ from the public key due to the elliptic curve discrete logarithm problem.

\textbf{Attack 3:} \textit{An attacker can perform the man-in-the-middle attack to the communication between the selected guard drone and other drone swarms.}

\textit{Prevention.} The attacker can interrupt the traffic, but the key agreement step is the prevention of the man-in-the-middle attacks.

\section{Conclusion}
In this study, the authentication of the new member to be included in the drone swarm and the authentication process between two different swarms of drones are examined. The authentication process in 5G NR standard and the proposed method in the study were tested with SimuLTE and the results were compared. According to the results obtained from SimuLTE, the proposed solution provides a time advantage compared to the solutions in the 5G NR standard.

\vspace{12pt}
\end{document}